\documentclass{ws-procs9x6-cpt22}

\begin{document}

\newcommand{\refeq}[1]{(\ref{#1})}
\def\etal {{\it et al.}}
%any other macros go here
\newcommand{\tcr}{\textcolor{red}}
\newcommand{\tcb}{\textcolor{blue}}
\newcommand{\beq}{\begin{equation}}
\newcommand{\eeq}{\end{equation}}
\newcommand{\ba}{\begin{eqnarray}}
\newcommand{\ea}{\end{eqnarray}}
\newcommand{\mbf}{\mathbf}
\newcommand{\gv}[1]{\ensuremath{\mbox{\boldmath$ #1 $}}} 

\title{Cherenkov radiation in chiral media}

\author{E.\ Barredo and L.F.\ Urrutia}
\address{High Energy Physics Department, Instituto de Ciencias Nucleares, Universidad Nacional Aut\'onoma de M\'exico,\\
Ciudad de M\'exico, 04510, M\'exico}

\begin{abstract}
 In the framework of Carrol-Field-Jackiw electrodynamics we calculate the spectral distribution of the Cherenkov radiation (CHR) produced by a charge moving at constant velocity in a chiral medium. We find zero, one or two Cherenkov angles according to the relation between the velocity of the particle and the refraction index of the medium.  
\end{abstract}

\bodymatter

\section{Introduction}
In 1989 Carroll, Field and Jackiw introduced a modifications of $3+1$ Maxwell electrodynamics (CFJ-ED) by adding a Chern-Simons Lagrangian density, which required the presence of a constant four-vector
$\tilde{b}_\mu=(b_0/c,\mathbf{b})$ \cite{Carroll1990},
\begin{equation}
\mathcal{L} = -\frac{1}{16\pi} F_{\mu\nu}F^{\mu\nu} -\frac{1}{c}J^\mu A_\mu+\frac{1}{8\pi}\tilde{b}_\mu \tilde{F}^{\mu\nu} A_\nu. 
\label{LAGDENS}
\end{equation}
Since the CFJ model breaks Lorentz invariance explicitly we recognize it as a CPT-odd contribution of the photon sector of the   SME \cite{Colladay1998}. Remarkably, CFJ-ED describes the electromagnetic response of Weyl semimetals (chiral media) in condensed matter physics \cite{BURKOV}. In the Brillouin zone, the LIV coefficients
$b_0$ and $\mathbf{b}$ account for the separation  of the corresponding  Weyl nodes in energy and momentum, respectively.\cite{PRL} CFJ-ED is also a particular case of axion ED which was previously shown to produce reversed CHR in topological insulators.\cite{OJLU}

\section{CHR  in chiral media }

Let us  consider a charge $q$ moving at constant velocity $v$ in the direction parallel to $\mathbf{b}=(0,0,b)$, from $-\xi$  to $+\xi $ with $\xi \rightarrow \infty$ at the end of the calculation. To determine the electromagnetic fields we
start from the modified Maxwell equations for CFJ electrodynamics in vacuum and  at the end of the calculation introduce the refraction index $n$ of the media. In the Lorentz gauge we have
\begin{equation}
\left( \eta^{\mu\nu} \partial^2-\epsilon^{\mu\nu\rho\sigma} {\tilde b}_\rho \partial_\sigma \right)A_\nu= \frac{4\pi}{c} J^\mu, \label{EQ2}
\end{equation}
which describes an infinite chiral vacuum defined by ${\tilde b}_\mu=(0,0,0,b), \, b>0$. The Green's function (GF) of the operator  in Eq. (\ref{EQ2}) is 
\begin{eqnarray}
&&G^{\mu \nu }(\mathbf{x},\mathbf{x^{\prime }};\omega ) =-\left[ (k_0
^{2}+\nabla ^{2})g^{\mu \nu }+ik_0 b\epsilon ^{\mu \nu 30}-b\epsilon
^{\mu \nu 3i}\partial _{i}+{\tilde{b}}^{\mu }{\tilde{b}}^{\nu }\right] f(%
\mathbf{x},\mathbf{x^{\prime }};\omega ),  \label{FG} \nonumber\\
&& f(\mathbf{x},\mathbf{x^{\prime }};\omega )=\frac{i}{8\pi }\int_{0}^{\infty } 
\frac{k_{\perp} \, dk_{\perp}}{k_\parallel} \,
J_{0}\Big(R_{\perp } k_\perp%
\Big)\sum_{\eta=\pm1 }\frac{\eta}{b} \frac{e^{i\sqrt{k_{\parallel }^{2}+\eta b k_{\parallel }}Z}%
}{\sqrt{k_{\parallel }^{2}+ \eta bk_{\parallel }}},
\end{eqnarray}%
Here we have $k_0=\omega/c \, $,  $k_\parallel=+\sqrt{k_0^2-k_\perp^2}\, $, $ Z=|x_3-x'_3|\,$, $\mathbf{R}=\mathbf{x}-\mathbf{x'}, \, R=|\mathbf{R}|$. The subindex $\perp$ in a vector indicates its projection on the plane perpendicular to the direction $x_3$. The function $f(\mathbf{x},\mathbf{x^{\prime }};\omega )$ is evaluated by the stationary phase method in the radiation zone. A  further approximation to first order in $bc/\omega $  is introduced to determine the stationary phase point. From now on the refraction index $n$ of the material is restored. The final result for the electromagnetic potential 
$A^\mu (\mathbf{x},\omega) = \int d^3 \mathbf{x}'\;G^{\mu}_{\;\; \nu}(\mathbf{x},\mathbf{x}';\omega) J^\nu(\mathbf{x}',\omega)$ is 
\begin{equation}
A^{\mu }(\mathbf{x},\omega )= \sum_{\eta=\pm 1 }{W}_{\eta
}{}^{\mu \nu }(\mathbf{\hat{n}})\; {\mathcal J}_{\nu }({}\mathbf{k}{}_{\eta },\,\omega
)\left( \frac{1}{r}e^{ik_0 C_\eta(\theta)r}\right),  \label{EMPOT1}
\end{equation}
where $W_\eta{}^{\mu\nu}(\mathbf{\hat{n}})$ and  ${\mathcal J}_{\nu }({}\mathbf{k}{}_{\eta },\,\omega
)$ are 
complicated functions of the observation angles $\mbf{\hat{n}}( \theta, \phi)$ that we do not write.   For the purpose of discussing the Cherenkov angles is it enough to consider in detail
\beq
\quad C_\eta(\theta)=
\sin^2 \theta+ \cos^2\theta \sqrt{1+ \eta \, \alpha(n) \sec\theta}, \qquad \alpha(n)= {b c}/(\omega n^2).
\eeq
After calculating $\mbf{E}$ and $\mbf{B}$ in the radiation zone we obtain the spectral distribution of the radiation
\ba
\frac{d^2E}{d\Omega d\omega} = \frac{n \omega^2q^2}{4\pi^2 c^3} \,  \sum_{\eta=\pm 1} \frac{%
\sin^2[\xi \, {\tilde \Xi}_\eta ]}{{ \tilde \Xi}^2_\eta}\, \mathcal{U}_{\eta}, \quad  \tilde{\Xi}_\eta(\theta)= \frac{\omega }{v }\Big(1-\frac{ n v }{c} \tilde{C}_{\eta }(\theta )\cos \theta \Big). &&
\label{SPECDIST}
\ea
Again, the function ${\cal U}_\eta$ is not relevant for our discussion. The main point in Eq. (\ref{SPECDIST}) is the property 
\begin{equation}
\lim_{\xi \rightarrow \infty}\frac{\sin[\xi \Xi_\eta ]}{\Xi_\eta}  = \pi \delta \left( \frac{\omega}{v}(1-\frac{nv}{c}\cos\theta C_\eta(\theta))\right), 
\end{equation}
yielding the condition $H_\eta(\theta)\equiv \cos\theta \, C_\eta(\omega,\theta)=c/nv$, which determines the Cherenkov angles as the intersection of the curves $H_\eta(\theta)$ with the horizontal lines $c/nv$ for a given material and particle velocity. 
\begin{figure}[h!]
\centering
%\includegraphics[width=4.5in, height = 2.8in]{H1new1.eps}
%\includegraphics[width=4in, height = 3.2in]{H1new1.eps}
%\includegraphics[scale=0.2]{grayscale1.eps} 
%\hspace{.3cm}
\includegraphics[scale=0.38]{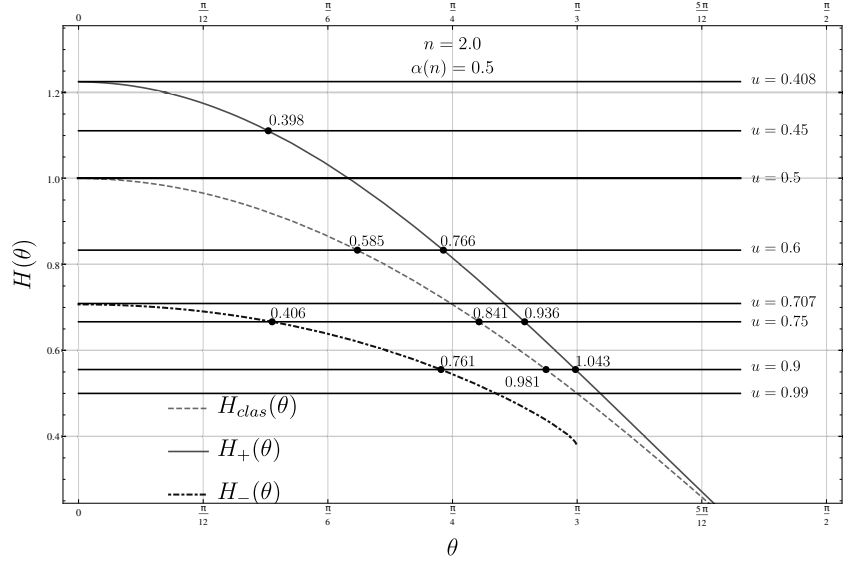}
\caption{Plot of the functions $H_{\rm clas}(\theta)$ (dashed line),   $H_+(\theta)$ (solid line) and $H_-(\theta)$ (dash-dotted line), for $\alpha(n)=0.5, \, n=2$. $H_{\rm clas}$ corresponds to $b=0$ and it is shown for comparison. The horizontal lines  labeled by $u=v/c$ correspond to $H=1/(nu)$  in the ordinate. }
\label{FIG1}
\end{figure}
Figure (\ref{FIG1}) shows the appearance of a different number of Cherenkov angles according to the particle velocity. The lower limit in the horizontal lines is $1/n$ corresponding to $u=1$. When $\alpha(n) < 1$, the curve $H_-(\theta)$ ranges over $0 < \theta < \arccos \alpha(n)$ and it is forbidden otherwise. A similar plot results in the case of vacuum CHR $(n=1, b\neq 0)$.\cite{PRL1}
\section*{Acknowledgments}
 The authors acknowledge support from the project CF/2019/428214.

\end{document}